\documentclass[12pt]{article}
\usepackage{graphicx}

\usepackage{bm} 

\newcommand\pubnumber{}
\newcommand\pubdate{\today}

\def\kobe{Department of Physics, Graduate School of Science, Kobe University,\\
Kobe, Hyogo 657-8501, Japan}
\def\kyoto{Department of Physics, Graduate School of Science, Kyoto University,\\
Kitashirakawa Oiwake-cho, Sakyo-ku, Kyoto 606-8502, Japan}

\def\Title#1{\begin{center} {\Large #1 } \end{center}}
\def\Author#1{\begin{center}{ \sc #1} \end{center}}
\def\Address#1{\begin{center}{ \it #1} \end{center}}

\newcommand\pubblock{\rightline{\begin{tabular}{l} \pubnumber\\
         \pubdate  \end{tabular}}}
\newenvironment{Abstract}{\begin{quotation}  }{\end{quotation}}
\newenvironment{Presented}{\begin{quotation} \begin{center} 
             PRESENTED AT\end{center}\bigskip 
      \begin{center}\begin{large}}{\end{large}\end{center} \end{quotation}}

\def\beq{\begin{equation}}
\def\eeq#1{\label{#1}\end{equation}}
\def\eeqn{\end{equation}}

\def\beqa{\begin{eqnarray}}
\def\eeqa#1{\label{#1}\end{eqnarray}}
\def\eeqan{\end{eqnarray}}

\let\bar=\overbar

\def\Dslash{\not{\hbox{\kern-4pt $D$}}}
\def\dslash{\not{\hbox{\kern-2pt $\del$}}}

\def\msb{{\bar{\ssstyle M \kern -1pt S}}}

\begin{document}
\begin{titlepage}
\pubblock

\vfill
\Title{Light Dark Matter Search with SOIPIX}
\vfill
\Author{Naoya Oka$^1$, Kentaro Miuchi$^1$, Takeshi G. Tsuru$^2$, Ayaki Takeda$^2$, and Hideaki Matsumura$^2$}
\Address{$^1$\kobe}
\Address{$^2$\kyoto}

\vfill
\begin{Abstract}
We propose a light dark matter search experiment using an SOI pixel detector (SOIPIX).  The event-driven SOIPIX can be a powerful tool for detecting light WIMPs because of its low energy threshold ($<$ 1 keV) and high timing resolution (few $\mu$s). In this study, we evaluate the performance of an SOIPIX prototype detector and we examine the required specifications of SOIPIX for our target sensitivity.

\end{Abstract}
\vfill
\begin{Presented}
International Workshop on SOI Pixel Detector (SOIPIX2015)\\ 
Tohoku University, Sendai, Japan, 3-6, June, 2015.
\end{Presented}
\vfill
\end{titlepage}
\def\thefootnote{\fnsymbol{footnote}}
\setcounter{footnote}{0}

\section{Introduction}

\begin{figure}[htb]
\centering
\includegraphics[height=8 cm]{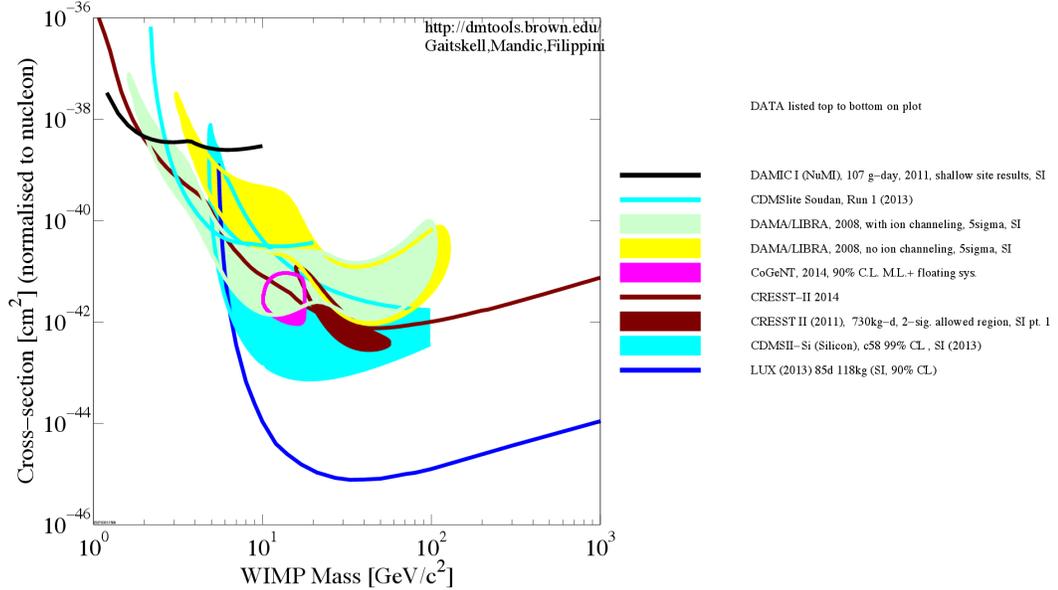}
\caption{Lines are showing upper limits with 90\% C.L. on WIMP-nucleon spin-independent cross sections from DAMIC \cite{Barreto:2011zu} (black),  SuperCDMS (CDMSlite) \cite{Agnese:2013jaa} (cyan), CRESST-II  \cite{Angloher:2014myn}(dark red), and LUX \cite{Akerib:2013tjd} (blue). Shaded areas correspond to regions of positive signals from DAMA/LIBRA interpreted by Savage et al. \cite{Savage:2008er} (light green: with ion channeling, 5 $\sigma$) (yellow: without ion channeling, 5$\sigma$), CoGeNT \cite{Aalseth:2014eft} (magenta, 90\% C.L.), CRESST-II \cite{Angloher:2011uu} (dark red, 2$\sigma$), and CDMS-II Si \cite{Agnese:2013rvf} (cyan, 99\% C.L.).  The CRESST experiment once reported a positive light WIMP signal, but they found no excess in recent upgraded data. CDMS-II reported excess using silicon data but this excess was not seen in germanium data.  Only DAMA/LIBRA and CoGeNT perform annual modulation analysis.
}
\label{fig:limit}
\end{figure}

The existence of dark matter is supported by various astronomical observations.  However, its nature remains unknown.  Weakly interacting massive particles (WIMPs) are promising candidates.  WIMPs can weakly couple with ordinary matters and can be directly detected using WIMP-nucleon scattering.  Among the various models of WIMP models, one that is well-motivated is supersymmetric neutralino.  Since the neutralino is expected to have a mass of over 50 GeV/$c^2$, most of experiments aim at this mass region.  Currently, the most stringent limit on WIMP-nucleon cross section has been set by the LUX experiment for mass above 7 GeV/$c^2$ \cite{Akerib:2013tjd}.  Meanwhile, DAMA experiment has been observed annual modulation of its signal since the year 1998\cite{Bernabei:2013xsa}, which can be interpreted as originating from WIMPs with masses of a few GeV/$c^2$ or a few tens of GeV/$c^2$.  Furthermore, other experiments, such as CoGeNT\cite{Aalseth:2014eft} and CDMS-Si\cite{Agnese:2013rvf} have recently seen positive signals in this region.  WIMPs in this range are called light WIMPs or low-mass WIMPs.  Since those signals are inconsistent with the result from LUX and other experiments, much discussion is taking place on this topic and is not yet concluded. Our experiment aims to search for light WIMPs using an SOI Pixel detector. Fig.\ref{fig:limit} shows the regions where DAMA and other experiments saw positive signals as well as the limitting curves by LUX and other experiments.

\section{Direct detection of light WIMPs and preceding study}
\begin{figure}[htb]
\centering
\includegraphics[height=8 cm]{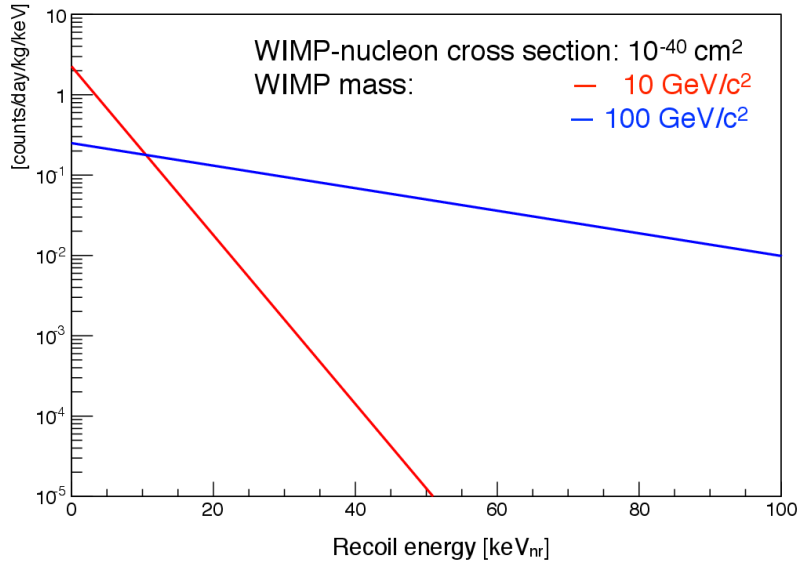}
\caption{Expected energy spectrum of silicon recoil by WIMPs with masses of 10 GeV/$c^2$ and 100 GeV/$c^2$.  The WIMP-nucleon cross section is set to $10^{-40}$ cm$^2$. Detector responses, such as acceptance or quenching effects, are not considered in this figure.}
\label{fig:Sirecoil}
\end{figure}

Velocity of WIMPs in our galaxy, $\bm{v}$, is considered to follow a Maxwell-Boltzmann distribution:
\[
f(\bm{v}) = e^{-\bm{v}^2/v_0^2}.
\]
The most probable speed of WIMPs $v_0$  is considered to be approximately 230 km/s \cite{Lewin:1995rx}. To detect WIMPs, we look for elastic (or sometimes inelastic) scattering of WIMP and nucleus.  In our case, silicon nuclei are the  targets for WIMPs.  Fig. \ref{fig:Sirecoil} shows the silicon's calculated recoil energy spectrum for WIMPs with masses of 10 GeV/c$^2$ and 100 GeV/c$^2$.  From this, we can see that, as the WIMP mass decreases, the fraction of lower energy events becomes larger.  Thus, a low energy threshold is the key for detection of light WIMPs.  Here, ``nr'' in keV$_{nr}$ stands for nuclear recoil, which can be by WIMPs, neutron, and so on.  Electron recoil, which can be by electrons, gammas, and so on is expressed with ``ee'' like keV$_{ee}$, which stands for electron-equivalent.  Because of an effect called ``quenching'', the observed energy of nuclear recoil with certain energy is smaller than that of electron recoil with the same energy.  Relative ionization yield by nuclear recoil compared to that of electron recoil is refered to as a quenching factor. Quenching factor depends on recoil energy and the detector material.  For silicon, it is about 0.2 at 1 keV$_{nr}$.  Currently, the DAMIC experiment has the world-lowest energy threshold among dark matter search experiments.  In 2011, using only 0.5 g of charge coupled devices (CCDs) with a 0.04 keV$_{ee}$ threshold, they have set a limit comparable with that of a kg-scale experiment \cite{Barreto:2011zu}.  Cosmogenic neutrons produced by cosmic-ray muons were their largest background (BG) contribution. Though those neutrons can be removed by active muon veto, this is not available for CCD because of its poor timing resolution (typically $\sim$ms).

\section{SOIPIX and its advantage}
A Silicon-On-Insulator (SOI) Pixel detector (SOIPIX) is being developed by KEK and other institutes \cite{Arai:2011ara}.  It is monolithic and comprises a thick, high-resistivity substrate (sensor part), a thin, low-resistivity silicon layer (CMOS circuit part), and a buried oxide (BOX) layer (SiO$_2$ insulator) between them.  In SOIPIX, each pixel has its own readout circuit. Thus, by extracting trigger signal from each pixel, we can read pixel by pixel (event-driven readout), unlike a CCD, which has to read from all pixels every time (frame readout).  This contributes a good timing resolution (few $\mu$s), which enables us to take advantage of anti-coincidence measurement.  Event-driven SOIPIX are being developed by X-ray astronomy group of Cosmic Ray group, Kyoto University and they are called XRPIX series \cite{Tsuru:2014jfa,Takeda:2014tza}. Currently, $35\, e^-$ readout noise (RMS) is achieved with the frame readout mode of an XRPIX3b sensor\cite{Takeda:JINST}.  Since the average energy for electron-hole creation in silicon is approximately 3.6 eV, $35\, e^-$ approximately corresponds to 0.1 keV.  If we set the energy threshold for the WIMP search to 5 times the readout noise, a 0.5 keV$_{ee}$ threshold is already achievable. Currently, the noise in event-driven readout mode is larger than that in frame readout mode, but this will be improved.  We can identify particles by tracks using pixel image from SOIPIX just like DAMIC's CCD. Furthermore, we can reject BG by anti-coincidence, which cannot be performed with CCD.  Because of these advantages, we plan to use SOIPIX for the light WIMP search.

\section{Performance test of the detector and simulation}
\begin{figure}[htb]
\centering
\includegraphics[height=8 cm]{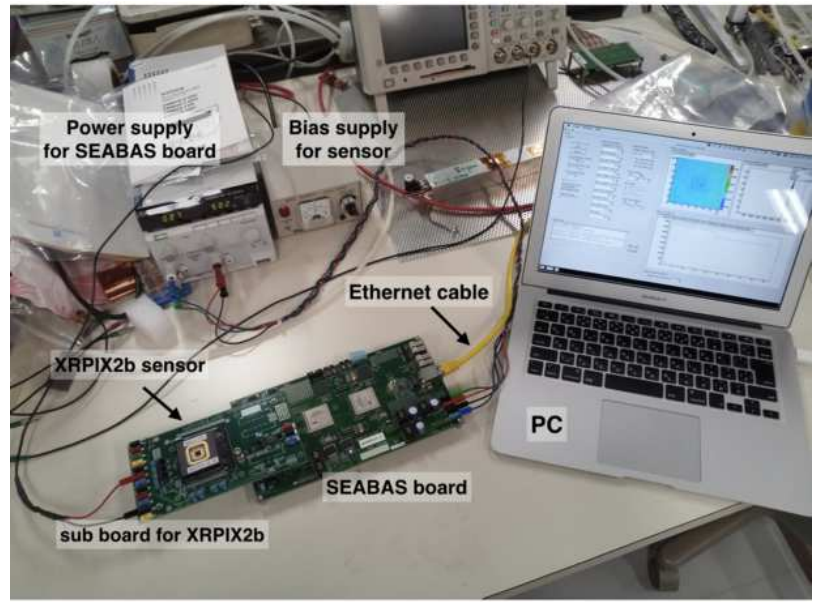}
\caption{The setup of our performance test.  An XRPIX2b sensor is placed on the readout circuit board.  The test could be done on top of the table.}
\label{fig:setup}
\end{figure}

We evaluated the performance of SOIPIX using XRPIX series detectors.  To date, tests at room temperature have been performed in the surface laboratory in Kobe university using the XRPIX2b sensor. For DAQ system, we used SEABAS (SOI Evaluation Board with SiTCP)\cite{Uchida:2008fha} that has two FPGA chips, one for controlling pixels and the other is for communicating with PC via Ethernet using a TCP/IP protocol.  Fig. \ref{fig:setup} shows our setup.

\begin{figure}[htb]
\centering
\includegraphics[height=8 cm]{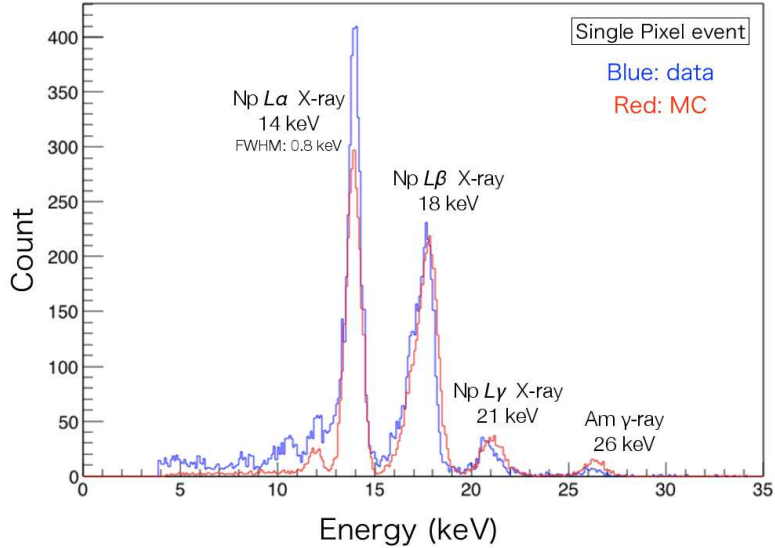}
\caption{Energy spectrum of single-pixel event data and MC of $^{241}$Am with the XRPIX2b sensor at room temperature. Energy calibration was done using 14 keV, 18 keV, 21 keV, and 26 keV peaks of $^{241}$Am.  Th energy threshold for each pixels was set to 4.5 keV.  The count rate of the MC is scaled to fit the 18 keV peak.}
\label{fig:spectrum}
\end{figure}

Calibration data were taken with $^{241}$Am in frame readout mode.  A simple detector Monte Carlo simulation (MC) based on Geant4 \cite{Agostinelli:2002hh, Allison:2006ve} was developed and its results were compared with $^{241}$Am data. Fig. \ref{fig:spectrum} shows the energy spectrum of single-pixel events of $^{241}$Am and its MC.  Though it is a simple one, this MC reproduces each peak of $^{241}$Am well.  We are going to improve the MC with more detailed geometries and processes.

\section{Estimation of sensitivity and the next step}

\begin{figure}[htb]
\centering
\includegraphics[height=9 cm]{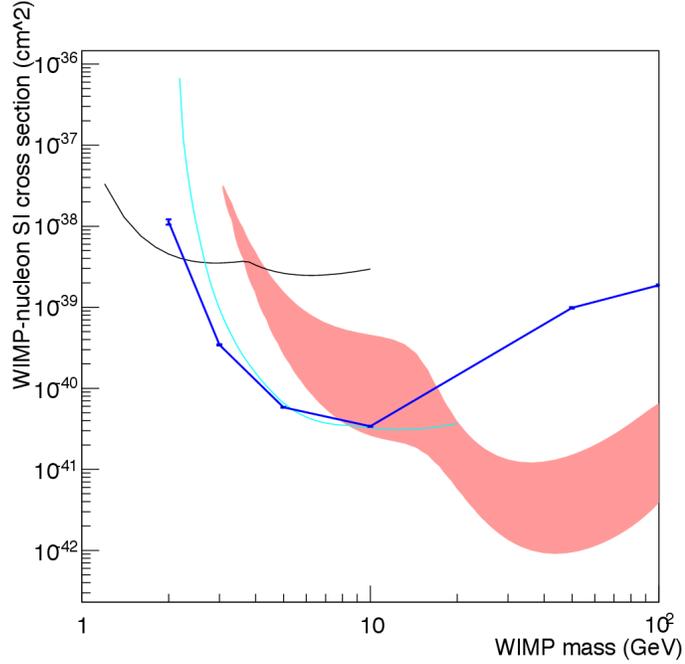}
\caption{Expected sensitivity to WIMPs with SOIPIX (shown in blue, 0.1 kg$\cdot$year) compared with other experiments; DAMA/LIBRA (pink), DAMIC (black), and SuperCDMS-CDMSlite (cyan). }
\label{fig:sensitivity}
\end{figure}

Using the MC explained in the previous section, we estimated the sensitivity of the SOIPIX against light WIMPs as shown in Fig. \ref{fig:sensitivity}. For this plot, the energy threshold was set to 0.3 keV$_{ee}$ by assuming a readout noise of 15 $e^-$, and the exposure was set to 0.1 kg$\cdot$year.  The BG level of 10 keV/kg/day was assumed with a 99\% background reduction by muon veto from DAMIC's neutron BG \cite{Barreto:2011zu} . A reasonable reduction value is assumed considereing preceding studies\cite{Bungau:2005xp}.  We plan to evaluate the performance of BG reduction by anti-coincidence measurement this year by using scintillators as veto detectors. Furthermore, radioactive contamination will be measured to obtain more detailed information on BG.  The quenching factor was simply set to 0.2 for the eintire energy range. We are planning to evaluate this factor by coincidence measurement of neutrons and gammas from $^{252}$Cf.  To achieve the target sensitivity, we are planning to develop a new SOIPIX for the WIMP search. Our target specification and that of the current XRPIX series are shown in table \ref{tab:spec}.

\begin{table}[h]
\begin{center}
\caption{XRPIX series and our target specification}
\begin{tabular}{c|cccc}  \hline
     Sensor         & Mass  & Readout noise \\ \hline
 current XRPIX  &   0.02 g (XRPIX2b) &   35 $e^-$ (XRPIX3b, -50 C$^\circ$, frame mode)\cite{Takeda:JINST}\\
 target specification	 &  1 g     &    $< 15 e^-$ \\ \hline
\end{tabular}
\label{tab:spec}
\end{center}
\end{table}

\section{Conclusion}
We  have commenced a light WIMP search experiment using SOIPIX.  A brief performance check of a prototype detector was undertaken and an MC simulation that included a detector response was developed.  Using that simulation, we estimated the sensitivity to light WIMPs and the SOIPIX specification required for that.

\pagebreak

\end{document}